\documentclass[fleqn,12pt,twoside,aastex]{article}
\usepackage{espcrc1,epsfig}

\usepackage{graphicx}
\usepackage[figuresright]{rotating}

\newcommand{\AmS}{{\protect\the\textfont2
  A\kern-.1667em\lower.5ex\hbox{M}\kern-.125emS}}

\title{The reactions and ashes of thermonuclear explosions on neutron stars}%
\author{J.L.~Fisker\address[ND]{Department of Physics and Joint Institute for Nuclear Astrophysics, University of Notre Dame, Notre Dame, IN 46556}\thanks{JLF is supported through the Joint Institute of Nuclear Astrophysics by NSF-PFC grant PHY02-16783.}, 
    E.F.~Brown\address[MSU1]{Department of Physics and Astronomy and Joint Institute for Nuclear Astrophysics, Michigan State University, East Lansing, MI 48824-2320},
    M.~Liebend\"orfer\address{CITA, University of Toronto, Toronto, Ontario M5S 3H8, Canada},
    F.-K.~Thielemann\address[Basel]{Department of Physics and Astronomy, Klingelbergstrasse 82, 4056 Basel, Switzerland}\thanks{FKT is supported by Swiss NSF grant 20-068031.02.}
    and M.~Wiescher\addressmark[ND]}
\begin{document}
\bibliographystyle{h-elsevier}

\maketitle

\begin{abstract}
This paper reports on the detailed $rp$-process reaction flow  on an accreting neutron star and the resulting ashes of a type I X-ray burst. It is obtained by coupling a 298 isotope reaction network to a self-consistent one-dimensional model calculation with a constant accretion rate of $\dot{M}=10^{17}\textrm{g}/\textrm{s}$.
\end{abstract}

\setcounter{footnote}{0}
\section{Introduction}
A neutron star in a low mass X-ray binary accretes a mixture of hydrogen and helium from the secondary star. 
The photospheric impact ionizes the matter completely afterwhich the matter spreads around the star, where it sinks down and develops a thermonuclear instability after a few hours or days \cite{Woosley76}. 
The instability causes an explosive runaway, where rapid and consequtive hydrogen and helium captures compete with $\beta^+$-decays to fusion the accreted matter into heavier proton-rich material \cite{Wallace81}.

The specific reaction flow depends on the ignition composition, the temperature and density evolution and the capture-rates, decay-rates and photodisintegration-rates of the nuclei in the reaction flow \cite{Koike99}.
However, near the dripline most of these rates are unknown which forces models to rely on theoretical predictions based on global models. 

In recent years new experimental developments -- in particular the introduction of radioactive beams -- have made it possible to measure these rates experimenally \cite{Schatz02}.  
Yet radioactive ion beam experiments take a long time to prepare and execute, so it is important to accurately establish the reaction flow to determine which reactions are critical in determining the reaction flow in the X-ray burst.

The charged particle reactions of the $rp$-process are strongly temperature-dependent. 
Therefore realistic values of the temperature evolution during the burst must be provided. 
Currently, this is done best by a self-consistent one-dimensional model.

\section{The thermonuclear runaway, the reaction flow, and the resulting ashes}
\begin{figure}[t!]
\begin{minipage}[t]{0.48\linewidth}
\includegraphics[width=\linewidth]{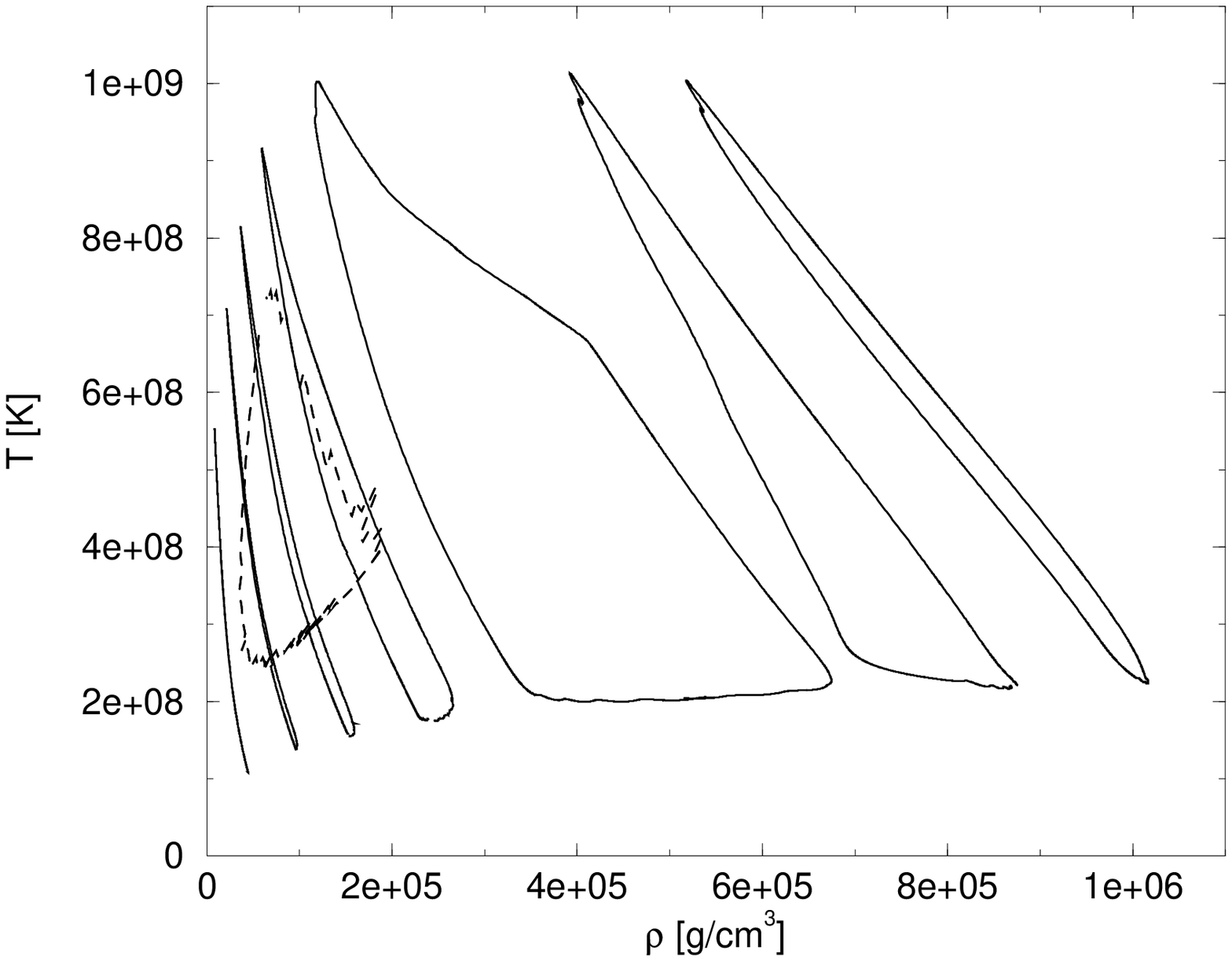}
\caption{From left to right (solid line): $y=2.1\times 10^6 \textrm{g}/\textrm{cm}^2$ (surface), $y=9.5\times 10^6 \textrm{g}/\textrm{cm}^2$, $y=1.9\times 10^7 \textrm{g}/\textrm{cm}^2$, $y=3.3\times 10^7 \textrm{g}/\textrm{cm}^2$, $y=6.2\times 10^7 \textrm{g}/\textrm{cm}^2$ (above ignition), $y=8.3\times 10^7 \textrm{g}/\textrm{cm}^2$ (ignition point), and $y=1.1\times 10^8 \textrm{g}/\textrm{cm}^2$ (ocean), where $y=\int_{R-r}^R\rho dr$ is the column depth. The dashed line indicate the region which is convective during the rising of the burst.}\label{fig:rhot} 
\end{minipage}\hfill
\begin{minipage}[t]{0.48\linewidth}
\includegraphics[width=\linewidth]{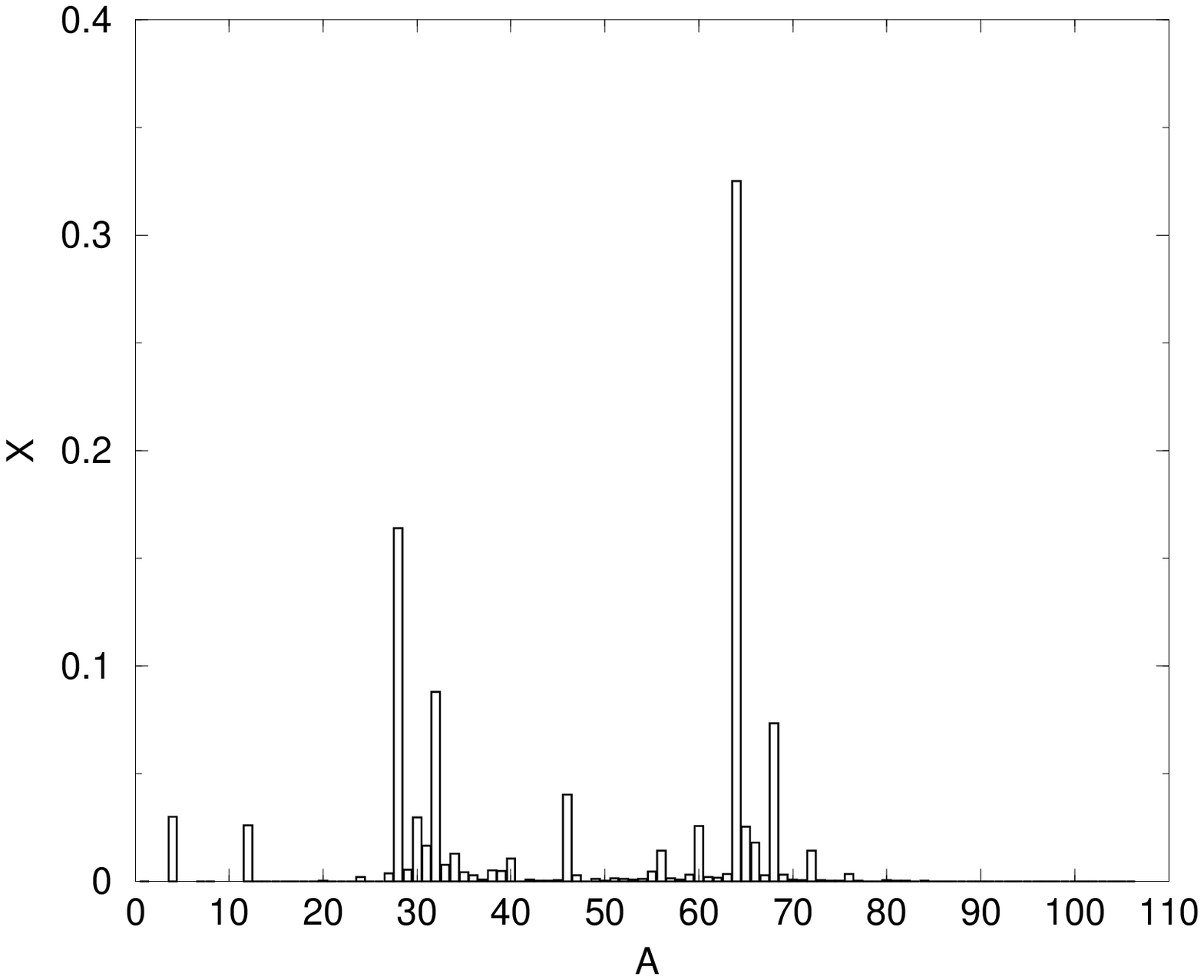}
 \caption{The mass fraction of the ashes about midway in between bursts plotted as a function of nucleon number: The peaks correspond to ${}^{4}\textrm{He}$; ${}^{12}\textrm{C}$; ${}^{28}\textrm{Si}$; ${}^{32}\textrm{S}$ and associated isotopes reacting in this region; ${}^{40}\textrm{Ca}$;  ${}^{46}\textrm{Ti}$;  and ${}^{56}\textrm{Ni}$, ${}^{60}\textrm{Zn}$, ${}^{64}\textrm{Ge}$, ${}^{68}\textrm{Se}$, ${}^{72}\textrm{Kr}$, and ${}^{76}\textrm{Sr}$ and their respective daughters.}\label{fig:oceanash} 
\end{minipage}
\end{figure}

This paper uses the same model as \cite{Fisker04b}. 
It considers a neutron star accreting at a constant rate of $\dot{M}=1\times 10^{17}\textrm{g}/\textrm{s}$ which corresponds to the inferred accretion rate of GS~1826-24 \cite{Galloway04}. 
Fig.~\ref{fig:rhot} shows the density and temperature for seven different depths as a function of time for a full cycle.\footnote{Densities and temperatures for different depths as a function of time along with initial compositions are available for post-processing calculations at http://www.nucastrodata.org}
The mixed H/He bursters have the same kind of reaction flow throughout the atmosphere; where the maximum temperature at a given depth generally determines how much material is processed. 
It also determines the extent of the flow, except for deeper layers, where the exhaustion of hydrogen limits the flow to heavier isotopes.

The composition is set by steady quiescent burning of freshly accreted hydrogen and residual hydrogen from the surface ashes from the previous burst steadily increasing the fraction of helium through the $\beta^+$-limited hot CNO cycle \cite{Woosley04}. 
At this temperature ($T\sim 0.2\textrm{GK}$ c.f.~Fig.~\ref{fig:rhot}) the ${}^{15}\textrm{O}(\alpha,\gamma){}^{19}\textrm{Ne}$-rate dominates the otherwise much faster ${}^{14}\textrm{O}(\alpha,p){}^{17}\textrm{F}$-rate \cite{Hahn96} and it is responsible for generating a distribution of seed nuclei up to the well-bound ${}^{40}\textrm{Ca}$ isotope for the $rp$-process which are important to the runaway \cite{Champagne92} that ensues at a column depth of $y=8.3\times 10^7 \textrm{g}/\textrm{cm}^2$. 

Initially the breakout of the hot CNO cycle is led by ${}^{15}\textrm{O}(\alpha,\gamma)$ ${}^{19}\textrm{Ne}(p,\gamma)$ ${}^{20}\textrm{Na}(p,\gamma)$ ${}^{21}\textrm{Mg}(p,\gamma)(\gamma,p)$ ${}^{22}\textrm{Al}$ which is blocked by photodisintegration, whence the flow proceeds via ${}^{21}\textrm{Mg}(\beta^+, T_{1/2}=0.122\textrm{s})$ ${}^{21}\textrm{Na}(p,\gamma)$ ${}^{22}\textrm{Mg}$. 
However, as the temperature rises and ${}^{15}\textrm{O}$, which depends on 
the $T_{1/2}=76.4\textrm{s}$ ${}^{14}\textrm{O}$ beta-decay, is depleted, the reaction flow follows ${}^4\textrm{He}(2\alpha,\gamma)$ ${}^{12}\textrm{C}(p,\gamma)$ ${}^{13}\textrm{N}(p,\gamma)$ ${}^{14}\textrm{O}(\alpha,p)$ ${}^{17}\textrm{F}(p,\gamma)$ ${}^{18}\textrm{Ne}$, which decays and returns via the hot CNO bicycle (see \cite{Wallace81}) until the temperature reaches $T\sim 0.9\textrm{GK}$ at which point the flow joins the other breakout with ${}^{18}\textrm{Ne}(\alpha,p)$ ${}^{21}\textrm{Na}(p,\gamma)$ ${}^{22}\textrm{Mg}$.

From ${}^{22}\textrm{Mg}$ two processes lead the flow toward heavier nuclei, namely the $(\alpha,p)$-process and the $rp$-process, whose different time-scales may lead to observable structure in the burst luminosity \cite{Fisker04b,Fisker04c}. 
The $(\alpha,p)$-reactions primarily occur on longer-lived nuclei in the $rp$-process flow, but since the cross section decreases as the Coulomb barrier increases due to the increasing charge of the target as the flow proceeds, these reactions are only effective for $A < 40$.

Higher temperatures make the proton-captures in the $rp$-process faster than the corresponding $\beta^+$-decays and move the reaction flow towards the proton dripline. Yet for temperatures above $\sim 1\textrm{GK}$ the tail of the Boltzmann distributed photons becomes sufficently energetic to photodisintegrate the more weakly bound proton-rich nuclei, so the flow moves away from the dripline \cite{Thielemann94}. 
If the flow branches into different paths, the branching ratio depends on the rate of the $\beta^+$-decay and the proton-capture, whence it becomes important to determine these rates accurately.

Here the $rp$-process flow branches into either ${}^{22}\textrm{Mg}(p,\gamma)$ ${}^{23}\textrm{Al}(p,\gamma)$ ${}^{24}\textrm{Si}(\beta^+, T_{1/2}=0.102\textrm{s})$ ${}^{24}\textrm{Al}$ or ${}^{22}\textrm{Mg}(\beta^+, T_{1/2}=3.86\textrm{s})$ ${}^{22}\textrm{Na}(p,\gamma)$ ${}^{23}\textrm{Mg}(p,\gamma)$ ${}^{24}\textrm{Al}$.
This is followed by ${}^{24}\textrm{Al}(p,\gamma)$ ${}^{25}\textrm{Si}$, which is in $(p,\gamma)(\gamma,p)$-equilibrium with ${}^{26}\textrm{P}$, where at high temperature, a fast additional proton capture creates the short-lived ${}^{27}\textrm{S}$, which decays to ${}^{27}\textrm{P}$. 
Otherwise the flow follows ${}^{25}\textrm{Si}(\beta^+, T_{1/2}=0.634\textrm{s})$ ${}^{25}\textrm{Al}(p,\gamma)$ ${}^{26}\textrm{Si}$, which may either decay or go to ${}^{27}\textrm{P}$ after an additional proton capture.
The flow continues with ${}^{26}\textrm{Si}(\beta^+, T_{1/2}=2.23\textrm{s})$ ${}^{26}\textrm{Al}(p,\gamma)$ ${}^{27}\textrm{Si}(p,\gamma)$ ${}^{28}\textrm{P}$ or ${}^{27}\textrm{P}(p,\gamma)$ ${}^{28}\textrm{S}(\beta^+, T_{1/2}=0.125\textrm{s})$ ${}^{28}\textrm{P}$ or ${}^{27}\textrm{P}(\beta^+, T_{1/2}=0.260\textrm{s})$ ${}^{27}\textrm{Si}(p,\gamma)$ ${}^{28}\textrm{P}$, which after a proton capture reaches ${}^{29}\textrm{S}(\beta^+, T_{1/2}=0.187\textrm{s})$ ${}^{29}\textrm{P}(p,\gamma)$ ${}^{30}\textrm{S}$.

The decay half-time of ${}^{30}\textrm{S}$ is rather long ($T_{1/2}=1.18\textrm{s}$), so initially the flow proceeds from ${}^{31}\textrm{Cl}$ which is in 
$(p,\gamma)(\gamma,p)$-equilibrium, since the $Q$-value is only $296\textrm{ keV}$, until the temperature becomes sufficiently high to completely photodisintegrate most of the ${}^{31}\textrm{Cl}$.
At this point the flow must proceed through the sulfur waiting point i.e.~${}^{30}\textrm{S}$ $(\beta^+, T_{1/2}=1.18\textrm{s})$ ${}^{30}\textrm{P}(p,\gamma)$ ${}^{31}\textrm{S}(p,\gamma)$ ${}^{32}\textrm{Cl}(\beta^+, T_{1/2}=0.298\textrm{s})$ ${}^{32}\textrm{S}(p,\gamma)$ ${}^{33}\textrm{Cl}(p,\gamma)$ ${}^{34}\textrm{Ar}$.
The proton-capture $Q$-value on ${}^{34}\textrm{Ar}$ is even lower, ($Q=78\textrm{ keV}$), so photodisintegration is stronger which makes ${}^{34}\textrm{Ar}$ another waiting point. 
Therefore the flow proceeds via ${}^{34}\textrm{Ar}(\beta^+, T_{1/2}=0.844\textrm{s})$ ${}^{34}\textrm{Cl}(p,\gamma)$ ${}^{35}\textrm{Ar}(p,\gamma)$ ${}^{36}\textrm{K}$.
Here the flow branches and proceeds either via ${}^{36}\textrm{K}(\beta^+, T_{1/2}=0.342\textrm{s})$ ${}^{36}\textrm{Ar}(p,\gamma)$ ${}^{37}\textrm{K}$ or ${}^{36}\textrm{K}(p,\gamma)$ ${}^{37}\textrm{Ca}(\beta^+, T_{1/2}=0.175\textrm{s})$ ${}^{37}\textrm{K}$, which captures a proton to ${}^{38}\textrm{Ca}$.
${}^{39}\textrm{Sc}$ is almost proton unbound, so the flow must wait for ${}^{38}\textrm{Ca}$ $(T_{1/2}=0.440\textrm{s})$ ${}^{38}\textrm{K}(p,\gamma)$ ${}^{39}\textrm{Ca}$.

Here the flow branches to either ${}^{39}\textrm{Ca}(p,\gamma)(\gamma,p)$ ${}^{40}\textrm{Sc}(p,\gamma)$ ${}^{41}\textrm{Ti}(\beta^+, T_{1/2}=0.080\textrm{s})$ ${}^{41}\textrm{Sc}(p,\gamma)$ ${}^{42}\textrm{Ti}$ or ${}^{39}\textrm{Ca}(\beta^+, T_{1/2}=0.860\textrm{s})$ ${}^{39}\textrm{K}(p,\gamma)$ ${}^{40}\textrm{Ca}(p,\gamma)$ ${}^{41}\textrm{Sc}(p,\gamma)$ ${}^{42}\textrm{Ti}$, where the latter path is active during the initial stage of the runaway burning the ${}^{40}\textrm{Ca}$ formed during the quiescent stage.
Now the flow branches to either ${}^{42}\textrm{Ti}(\beta^+, T_{1/2}=0.199\textrm{s})$ ${}^{42}\textrm{Sc}(p,\gamma)$ ${}^{43}\textrm{Ti}(p,\gamma)$ ${}^{44}\textrm{V}$ or ${}^{42}\textrm{Ti}(p,\gamma)(\gamma,p)$ ${}^{43}\textrm{V}(p,\alpha)$ ${}^{44}\textrm{Cr}(\beta^+, T_{1/2}=0.053\textrm{s})$ ${}^{44}\textrm{V}$, where they connect and continue with ${}^{44}\textrm{V}(p,\gamma)$ ${}^{45}\textrm{Cr}(\beta^+, T_{1/2}=0.050\textrm{s})$ ${}^{45}\textrm{V}(p,\gamma)$ ${}^{46}\textrm{Cr}$, which decays to ${}^{46}\textrm{Ti}$ via ${}^{46}\textrm{V}$.

The flow proceeds to ${}^{49}\textrm{Mn}$ via ${}^{46}\textrm{Ti}(p,\gamma)$ ${}^{47}\textrm{V}(p,\gamma)$ ${}^{48}\textrm{Cr}(p,\gamma)$ ${}^{49}\textrm{Mn}$ or ${}^{46}\textrm{Cr}(p,\gamma)$ ${}^{47}\textrm{Mn}(p,\gamma)$ ${}^{48}\textrm{Fe}(\beta^+, T_{1/2}=0.048\textrm{s})$ ${}^{48}\textrm{Mn}$, where it meets ${}^{46}\textrm{V}(p,\gamma)$ ${}^{47}\textrm{Cr}(p,\gamma)$ ${}^{48}\textrm{Mn}(p,\gamma)$ ${}^{49}\textrm{Fe}(\beta^+, T_{1/2}=0.070\textrm{s})$ ${}^{49}\textrm{Mn}$. 
At this branching point the proton capture is weaker than the $\beta^+$-decay! 
Therefore the main flow proceeds via ${}^{49}\textrm{Mn}(\beta^+ T_{1/2}=0.382\textrm{s})$ ${}^{49}\textrm{Cr}(p,\gamma)$ ${}^{50}\textrm{Mn}(p,\gamma)$ ${}^{51}\textrm{Fe}(p,\gamma)$ ${}^{52}\textrm{Co}(p,\gamma)$ ${}^{53}\textrm{Ni}$.

At this point the flow becomes more dependent on the nuclear masses, which determine the proton-capture $Q$-values, since the photodisintegration of the compound nucleus determines the specific waiting points. 
The waiting points are: ${}^{53}\textrm{Ni}$, ${}^{54}\textrm{Ni}$, and ${}^{55}\textrm{Ni}$, where the flow either proceeds with two fast proton captures to ${}^{57}\textrm{Zn}$ or awaits the $T_{1/2}=0.212\textrm{s}$ decay.
Similiarly ${}^{57}\textrm{Zn}$, ${}^{58}\textrm{Zn}$, ${}^{59}\textrm{Zn}$, and ${}^{60}\textrm{Zn}$ are waiting points. 
However, ${}^{60}\textrm{Zn}$ has a halflife of 2.38 minutes, so the flow must proceed with two proton captures reaching another row of waiting points: ${}^{62}\textrm{Ge}$, ${}^{63}\textrm{Ge}$, and ${}^{64}\textrm{Ge}$ which lives for 2.67 minutes. 
A proton-capture $Q$-value of $80\textrm{keV}$ essentially stops the reaction flow here, which most slowly decay out of ${}^{64}\textrm{Ge}$ to form heavier nuclei. 
This means that most burst ashes are concentrated on the ${}^{64}\textrm{Ge}$ daughters as can be seen from fig.~\ref{fig:oceanash}. 
Further waiting points which are reached during the $\sim 150\textrm{s}$ cooling time of the burst are ${}^{67}\textrm{Se}$, ${}^{68}\textrm{Se}$, ${}^{72}\textrm{Kr}$, and ${}^{76}\textrm{Sr}$ after which the flow becomes negligible.

Ashes heavier than $A=40$ are primarily made above the ignition point during a burst and subsequently accreted into the ocean, whereas lighter ashes are made directly in the ocean during the helium burning in the heat bath following a burst (see fig.~\ref{fig:oceanash}).

\section{Conclusion}
In this paper we have determined the $rp$-process reaction flow on an accreting neutron star using a self-consistent one-dimensional model. Waiting points and important reactions have been identified.
The ashes are concentrated around $A\sim 30$ and $A=64$ and show residual amounts of about 3\% ${}^{4}\textrm{He}$ and 3\% ${}^{12}\textrm{C}$.
This disagrees with earlier calculations of one-zone models which predicted much heavier ashes but lacked a self-consistent prescription for the heat transport. The present result corroborates other one-dimensinal models \cite{Woosley04}.

\end{document}